\documentclass[aps,prx,twocolumn,showpacs,superscriptaddress,groupedaddress,floatfix,10pt,amsmath,amssymb,footinbib]{revtex4-1}
\usepackage{graphicx}
\usepackage{bm,bbm}
\usepackage[urlcolor=blue]{hyperref}
\hypersetup{colorlinks=true,allcolors=blue}
\usepackage{amssymb} 

\newcommand{\Elr}[1]{\left\langle #1\right\rangle}
\newcommand{\E}[1]{\langle #1\rangle}

\newcommand{\rmd}{d}
\newcommand{\rme}{\mathrm{e}}

\newcommand{\f}[1]{\mathbf{#1}}
\newcommand{\x}{\f x}

\newcommand{\norm}[1]{\left\lvert\left\lvert #1 \right\rvert\right\rvert}

\newcommand{\bsig}{\boldsymbol{\sigma}}

\newcommand{\beps}{{\boldsymbol{\varepsilon}}}

\newcommand{\R}{\mathbb{R}}

\newcommand{\var}{{\rm var}}

\newcommand{\cov}{{\rm cov}}

\newcommand{\nxy}{\hat{n}_{xy}}

\newcommand{\nij}{\hat{n}_{ij}}
\newcommand{\rxy}{r_{xy}}\newcommand{\ryx}{r_{yx}}

\newcommand{\rij}{r_{ij}}
\newcommand{\tx}{\hat{\tau}_{x}}

\newcommand{\ti}{\hat{\tau}_{i}}
\newcommand{\sumxy}{\sum_{x\ne y}}

\newcommand{\sumij}{\sum_{i\ne j}}

\usepackage{color}\usepackage{xcolor}
\colorlet{mylinkcolor}{blue!66!black!80}
\definecolor{grey}{rgb}{0.6,0.6,.6}
\definecolor{darkgrey}{rgb}{0.4,0.4,.4}
\definecolor{darkgreen}{rgb}{0,0.4,0}
\definecolor{lightgreen}{rgb}{0,0.7,0}
\definecolor{darkred}{rgb}{0.5,0,0}

\newcommand{\blue}[1]{{\color{mylinkcolor}#1}}

\renewcommand{\blue}[1]{{#1}}

\begin{document}
\title{Direct Route
  to Thermodynamic Uncertainty Relations and Their Saturation}
\author{Cai Dieball}
\author{Alja\v{z} Godec}
\email{agodec@mpinat.mpg.de}
\affiliation{Mathematical bioPhysics Group, Max Planck Institute for Multidisciplinary Sciences, Am Fa\ss berg 11, 37077 G\"ottingen}
\begin{abstract}
Thermodynamic uncertainty relations (TURs) bound the dissipation in non-equilibrium systems from below by fluctuations of an
observed current. Contrasting the elaborate techniques
employed in existing proofs, we here prove 
TURs directly from the Langevin equation.~This establishes
the TUR as an inherent property of overdamped stochastic
equations of motion.~\blue{In addition, we extend the
  transient TUR 
  to
  currents and densities with explicit time-dependence.}~By including current-density correlations we,
moreover, derive a new
sharpened TUR for transient dynamics.~Our arguably simplest and most direct
proof\blue{, together with the new generalizations,} allows us to systematically determine conditions under which
the different TURs saturate and thus allows for a more accurate
thermodynamic inference.\ 
\blue{Finally we outline 
  the direct proof also for Markov jump dynamics.}
\end{abstract}
\maketitle
A defining characteristic of non-equilibrium systems is a
non-vanishing entropy production
\cite{Roldan2010PRL,Pigolotti2017PRL,Seifert2012RPP,Seifert2005PRL,Esposito2010PRE,VandenBroeck2010PRE,Vaikuntanathan2009EEL,Qian2013JMP}
emerging during relaxation
\cite{Vaikuntanathan2009EEL,Maes2011PRL,Qian2013JMP,Maes2017PRL,Shiraishi2019PRL,Lapolla2020PRL},
in the presence of time-dependent (e.g.\ periodic
\cite{Proesmans2017,Koyuk2018JPAMT,Barato2018NJP,Barato2019,Koyuk2019PRL,Koyuk2020PRL})
driving, or in non-equilibrium steady states (NESS)
\cite{Jiang2004,Maes2008PA,Maes2008EEL,Seifert2010EEL,Barato2015PRL,Gingrich2016PRL,Dieball2022PRL,Dieball2022PRR}. A
detailed understanding of the thermodynamics of systems far from
equilibrium is in particular required for unraveling the physical
principles that sustain active, living matter
\cite{Active,Active_1,Active_2,Active_3,Active_4}.
Notwithstanding its importance, the entropy production within a
non-equilibrium system beyond the linear response is virtually impossible to quantify from
experimental observations, as it requires detailed knowledge about all
dissipative degrees of freedom. 

A recent and arguably the most relevant method to infer a lower bound on the
entropy production in an experimentally observed complex system is via the so-called
thermodynamic uncertainty relation (TUR)
\cite{Horowitz2019NP,Gingrich2017JPAMT,Vu2020PRE,Manikandan2020PRL,Otsubo2020PRE,Li2019NC,Koyuk2021JPAMT,Dieball2022PRL,Dieball2022PRR,Dechant2021PRR},
which relates the (time-accumulated) dissipation $\Sigma_t$ to
fluctuations of a general time-integrated current $J_t$. For overdamped systems in a NESS it
reads \cite{Barato2015PRL,Gingrich2016PRL}
\begin{align}
  \frac{\Sigma_t}{k_{\rm B}T} \ge 2 \frac{\E{J_t}^2}{\var(J_t)}\,,
\label{TUR NESS} 
\end{align}
with variance ${\rm var}(J_t)\equiv\E{J_t^2}-\E{J_t}^2$ and thermal energy $k_{\rm B}T$, which will henceforth be
dropped for convenience and replaced by the convention of energies measured in
units of  $k_{\rm B}T$. The TUR may be seen as the natural counterpart
of the fluctuation-dissipation theorem \cite{Fu2022} or a more
precise formulation of the second law \cite{Dechant2021PRX}. Notably,
it may also
be interpreted as gauging the ``thermodynamic cost of precision''
\cite{cost}, and it was found to limit the temporal extent of
anomalous diffusion \cite{Hartich2021PRL}. 

Since its original discovery \cite{Barato2015PRL} and proof
\cite{Gingrich2016PRL} for systems in a NESS, a large number of more
or less general variants of the TUR were derived. In particular, for
paradigmatic overdamped dynamics and Markov jump processes, such
generalized TURs have been found for transient systems
\blue{(i.e.\ non-stationary dynamics emerging e.g.\ from non-steady-state initial conditions)} in absence
\cite{Pietzonka2017PRE,Dechant2018JSMTE,Liu2020PRL} and presence of
time-dependent driving \cite{Koyuk2019PRL,Koyuk2020PRL}.\ 
Moreover, an extension to state variables (which we will refer to as
``densities'') instead of currents has been formulated
\cite{Koyuk2020PRL}, and recently correlations of densities and
currents have been incorporated to significantly sharpen and even
saturate  the inequality for steady-state systems
\cite{Dechant2021PRX}. Note, however, that the validity of the TUR is generally
limited to overdamped dynamics, as it was 
shown to break down
in systems with momenta \cite{Partick_break}.

Many different techniques have been employed to derive TURs, including
large deviation theory
\cite{Gingrich2016PRL,Pietzonka2016PRE,Horowitz2017PRE,Gingrich2017JPAMT,Fu2022},
bounds to the scaled cumulant generating function
\cite{Dechant2018JSMTE,Dechant2020PNASU,Koyuk2020PRL},
as well as martingale \cite{Pigolotti2017PRL} and Hilbert-space
\cite{Falasco2020NJP} techniques.\ Most notably, the TUR has been
derived  as a consequence of the 
generalized Cram{\'e}r-Rao inequality
\cite{Dechant2018JPAMT,Liu2020PRL} which is well known in information
theory and statistics. However,
whilst
providing
valuable insight, the proof via the Cram{\'e}r-Rao inequality
includes quantifying the Fisher information of the Onsager-Machlup
path measure \cite{Dechant2018JPAMT} and 
involves a dummy parameter that 'tilts' the original dynamics. Thus,
it may not be faithfully considered as being direct. In fact,
the TUR and its generalizations seem to be
an inherent property of overdamped stochastic
dynamics and are thus, akin to quantum-mechanical uncertainty, expected to follow directly from the equations of motion.    

Here we show that no elaborated concepts beyond the equations of
motion are indeed required. Using only stochastic calculus
and the well known Cauchy-Schwarz inequality we
prove various existing TURs (including the correlation-TUR \cite{Dechant2021PRX}) for time-homogeneous overdamped dynamics in
continuous space directly from the Langevin equation. Thereby we both, unify and simplify,
proofs of TURs. Moreover, we derive, for the first time, the sharper correlation-TUR for
transient dynamics without explicit time-dependence. This improved TUR can be saturated arbitrarily far from equilibrium for any
initial condition and duration of trajectories\blue{, which we
  illustrate with the example of a displaced harmonic trap.}
Our simple
proof offers several advantages 
and we therefore believe 
that it deserves attention even in cases that have already been
proven before. Most notably it enables immediate insight into how one
can saturate the various TURs and allows for easy
generalizations. \blue{Beyond the results for overdamped dynamics, we
  illustrate the analogous direct proof of the steady-state
  TUR also for Markov jump dynamics.} 

\emph{Setup.---}\blue{We consider \blue{$d$-}dimensional \footnote{We
  consider $\R^d$ or a finite subspace with periodic or reflecting
    boundary conditions.}
  time-homogeneous (i.e.\ coefficients do not explicitly depend on time) overdamped dynamics described by the stochastic differential (Langevin) equation  \cite{Gardiner1985,Pavliotis2014TiAM}
\begin{align}
\rmd\x_\tau=\f F(\x_\tau)\rmd\tau+\bsig(\x_\tau)\circledast\rmd\f W_\tau\,,
\label{SDE} 
\end{align}
where the anti-It\^o product $\circledast$ assures thermodynamical
consistency in the case of multiplicative noise (i.e.\ space-dependent
$\bsig(\x_\tau)$)
\cite{Hartich2021PRX,Dieball2022PRR,Pigolotti2017PRL,anti,anti2}.~The
choice of the product is irrelevant in the case of additive noise $\bsig(\x_\tau)=\bsig$.~The increment $\rmd\f W_\tau$ of the Wiener process, has zero mean $\langle\rmd\f
W_\tau\rangle=\f 0$ and is due to its covariance 
$\langle\rmd W_{\tau,i}\rmd
W_{\tau',j}\rangle=\delta(\tau-\tau')\delta_{ij}\rmd\tau\rmd\tau'$ known as delta-correlated or white noise.} 
The noise
amplitude is related to the diffusion coefficient via $\f D(\x)\equiv
\bsig(\x)\bsig(\x)^T/2$ \blue{where $\bsig$ and $\f D$ are $d\times d$ matrices}.\ Let $P(\x,\tau)$ be the probability density
to find $\x_\tau$ at a point $\x$ given some initial condition $P(\x,0)$.\ Then the instantaneous probability density current $\f j(\x,\tau)$ is given by
\begin{align}
  \f j(\x,\tau)&=[\f F(\x)-\f D(\x)\nabla]P(\x,\tau)\,,
  \label{def probability current}
\end{align}
and the Fokker-Planck equation \cite{Risken1996,Pavliotis2014TiAM} for
the time-evolution of $P(\x,\tau)$ follows from Eq.~\eqref{SDE}
and reads \cite{Gardiner1985}
\begin{align}
\partial_\tau P(\x,\tau)=-\nabla\cdot\f j(\x,\tau)\,.\label{FPE} 
\end{align}
In the special case that $\f F(\x)$ is sufficiently confining a NESS is eventually reached
with invariant density $P_{\rm s}(\x)\equiv P(\x,\tau\to\infty)$ and steady-state
current $\f j_{\rm s}(\x)\equiv [\f F(\x)-\f D(\x)\nabla] P_{\rm
  s}(\x)$ with $\nabla\cdot\f j_{\rm s}(\x)=0$ \cite{Pavliotis2014TiAM}.
The mean total (medium plus system) entropy production in the time
interval $[0,t]$ is given by \cite{Seifert2005PRL,Seifert2012RPP}
\begin{align}
\Sigma_t&=\int\rmd\x\int_{0}^{t}\frac{\f j^T(\x,\tau)\f D^{-1}(\x)\f j(\x,\tau)}{P(\x,\tau)}d\tau
\,.\label{dissipation}
\end{align}
Let $J_t$ be a generalized time-integrated current
with some vector-valued $\f U(\x,\tau)$ defined via the Stratonovich \blue{stochastic} integral
\blue{(only for $\x$-dependent $\f U$ the convention matters)}
\begin{align}
J_t\equiv\int_{\tau=0}^{\tau=t}\f U(\x_\tau,\tau)\cdot\circ\rmd\x_\tau\,.
\label{def current Strato} 
\end{align}
Note that for any integrand $\f U$ this current and its first two
moments are readily obtained from measured trajectories
$(\x_\tau)_{0\le\tau\le t}$. Therefore a TUR involving such $J_t$ is ``operationally accessible''. For
dynamics  in Eq.~\eqref{SDE}
the current may be equivalently written as the sum of It\^o- and
$\rmd\tau$-integrals, $J_t=J_t^{\rm I}+J_t^{\rm II}$, with \cite{Dieball2022PRR}
\begin{align}
J_t^{\rm I}&\equiv\int_{\tau=0}^{\tau=t}\f U(\x_\tau,\tau)\cdot\bsig(\x_\tau)\rmd\f W_\tau\nonumber\\
J_t^{\rm II}&\equiv\int_0^t\big[\f U(\x_\tau,\tau)\cdot\f
F(\x_\tau)+\nabla\cdot\left[\f D(\x_\tau)\f U(\x_\tau,\tau)\right] \big]\rmd\tau\nonumber\\
&\equiv \int_0^t \mathcal U(\x_\tau,\tau)\rmd\tau\,.
\label{def current Ito}
\end{align}
By the zero-mean and independence properties of the Wiener process $\E{J_t^{\rm I}}=0$ and thus
$\E{J_t}=\E{J_t^{\rm II}}=\int_0^t\rmd\tau\int\rmd\x\mathcal
U(\x,\tau)P(\x,\tau)$. Integrating 
by parts and using
Eq.~\eqref{def probability current} we obtain (see also \cite{Dieball2022PRR})
\begin{align}
\E{J_t}=\int_0^t\rmd\tau\int\rmd\x\f U(\x,\tau)\cdot\f j(\x,\tau)\,.
\label{mean J} 
\end{align}
The 
variance $\var(J_t)$ can in turn be computed from two-point densities 
\cite{Dieball2022PRL,Dieball2022PRR,Dieball_ARXIV_JPhysA_1,Dieball2022JPA},
but is not required 
to prove TURs.

We now outline our direct proof of TURs. First, we re-derive the
classical TUR~\eqref{TUR NESS} and its generalization to
transients \cite{Dechant2018JSMTE}, whereby we find a novel correction
term that extends the validity of the transient TUR. Next we prove the
TUR for densities \cite{Koyuk2020PRL} and thereafter the
correlation-improved TUR \cite{Dechant2021PRX}, for the first time also
for non-stationary dynamics. Finally, we explain how to saturate the
various TURs \blue{and illustrate our findings with an example}.~The proof relies
solely on the equation of motion Eq.~\eqref{SDE} and implied
Fokker-Planck equation \eqref{FPE}, which is why we call the
proof ``direct''.  

\emph{Direct proof of TURs.---}\blue{The essence of the direct proof is fully contained in the following Eqs.~\eqref{def At}-\eqref{TUR from ansatz}.}
First, we require 
a scalar
quantity $A_t$ with zero mean and whose second moment yields
the dissipation defined in Eq.~\eqref{dissipation}, i.e.\ $\langle
A_t^2\rangle=\Sigma_t/2$ \footnote{The factor ``$1/2$'' is introduced for convenience}.\
Considering the
``delta-correlated'' 
property of $\rmd\f W_\tau$ and ${\f D=\f D^T=\bsig(\x)\bsig(\x)^T/2}$
leads to the ``educated guess'' \textcolor{black}{(see \cite{Note3})}
\begin{align}
A_t\equiv\int_{\tau=0}^{\tau=t}\frac{\f j(\x_\tau,\tau)}{P(\x_\tau,\tau)}\cdot[2\f D(\x_\tau)]^{-1}\bsig(\x_\tau)\rmd\f W_\tau\,,\label{def At} 
\end{align}
where $A_t$ cannot be inferred from trajectories since 
only $\rmd\x_\tau$ but not $\rmd\f W_\tau$ is observed.\ 
\blue{$A_t$ can be understood as the ``purely random'' part $\bsig(\x_\tau)\rmd\f W_\tau$ of the increment $\rmd\x_\tau$ weighted by the local velocity and inverse diffusion coefficient.}
Because $\langle
A_t J^{\rm I}_t\rangle=\langle J_t\rangle$ and
$\E{A_t\E{J_t}}=\E{A_t}\E{J_t}=0$ we have
\begin{align}
\E{A_t(J_t-\E{J_t})}=\E{J_t}+\E{A_tJ_t^{\rm II}}\,,\label{proof ansatz}
\end{align}
and the Cauchy-Schwarz inequality $\E{A_t(J_t-\E{J_t})}^2\le \E{A_t^2}\var(J_t)$
further yields 
\begin{align}
\frac{\Sigma_t}{2}\var(J_t)\ge\left[\E{J_t}+\E{A_tJ_t^{\rm II}}\right]^2\,.\label{TUR from ansatz}
\end{align}
Compared to Eq.~\eqref{proof ansatz}
the inequality \eqref{TUR from ansatz} has the advantage 
that $\var(J_t)$ is
operationally accessible and $\Sigma_t$ (unlike $A_t$) has a clear
physical interpretation.

To obtain the TUR we are left with evaluating $\E{A_tJ_t^{\rm II}}$, which involves the two-time correlation of $\rmd\f W_\tau$ and $\rmd\tau'$
integrals in Eq.~\eqref{def At} and Eq.~\eqref{def current Ito},
respectively. For times $\tau\ge\tau'$ this correlation vanishes due
to the independence property of the Wiener process. However,
non-trivial correlations occur for $\tau<\tau'$ because the
probability density of $\x_{\tau'}$ depends on $\rmd\f W_\tau$.
\blue{We quantify these correlations including $\rmd\f W_\tau$ by writing $\E{A_tJ_t^{\rm II}}$ as an average over the joint density to be at points $\x,\x+d\x,\x'$ at times $\tau<\tau+d\tau<\tau'$, respectively, and expanding 
\begin{align}
& P(\x',\tau'|\x+d\x,\tau+d\tau) \nonumber\\& 
=P(\x',\tau'|\x,\tau) + \rmd\x\cdot\nabla_\x P(\x',\tau'|\x,\tau) + \mathcal{O}(d\tau)\,.
\label{auxd}
\end{align}
Following this approach \cite{Dieball2022PRR,Dieball2022JPA} (or alternatively via Doob conditioning
\cite{Doob,Chetrite2014AHP,Pigolotti2017PRL} as in Ref.~\cite{Dechant2021PRR}) one can formulate a general calculation rule that in this case reads (for details see \cite{Note3})
\begin{align}
\E{A_tJ_t^{\rm II}}
=&-\int_0^t\rmd\tau'\int\rmd\x' \mathcal U(\x',\tau')\int_0^{\tau'}\rmd\tau\int\rmd\x\times\nonumber\\&\quad
P(\x',\tau'|\x,\tau)\nabla_\x\cdot\f j(\x,\tau)\,.
\label{lemma}
\end{align}}For steady-state systems 
we have $\nabla \cdot\f j(\x,\tau)=\nabla \cdot\f j_{\rm s}(\x)=0$
and thus $\E{A_tJ_t^{\rm II}}=0$, such that Eq.~\eqref{TUR from ansatz} immediately implies the original TUR in Eq.~\eqref{TUR NESS}.

\begin{figure*}[ht!!]
\begin{center}
\includegraphics[width=1.0\textwidth]{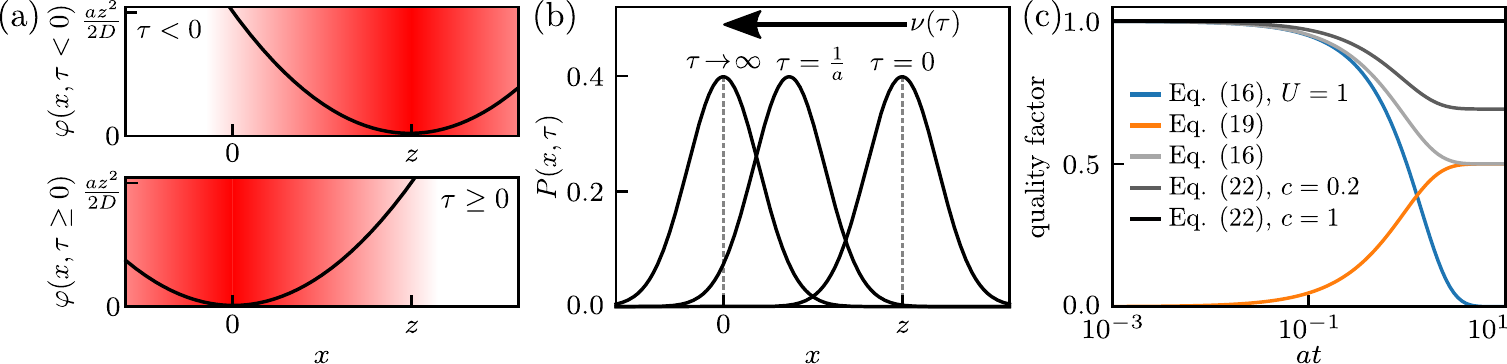}
\caption{\blue{(a)~Brownian particle in a one-dimensional harmonic
    trap with stiffness $a$,
    $\varphi(x,\tau)=a(x-x^0_\tau)^2/2D$ displaced from $x^0_{\tau<0}=z$
  to $x^0_{\tau\ge 0}=0$. Upon being initially equilibrated in
  $\varphi(x,\tau<0)=a(x-z)^2/2D$ (i.e.\ from the initial condition
  $p_0(x)\propto\exp[-a(x-z)^2/2D]$)
  the particle evolves 
  for $\tau\ge0$ due to $D\partial_x\varphi(x,\tau\ge0)=ax$
   according to
  $\rmd x_\tau=-ax_\tau d\tau+\sqrt{2D}\rmd W_\tau$ towards an equilibrium $p_{\tau\to\infty}(x)\propto\exp(-ax^2/2D)$. 
  (b)~Illustration of the evolution of $P(x,\tau)$ for $z=5\sqrt{D/a}$.~(c)~Quality factors defined as the ratio of right- and left-hand side of the
TURs as a function of the dimensionless quantity $at$.
  All
  quality factors turn out to be independent of $z,D$ and only depend
  on $a,t$ through $at$; explicit analytic expressions are given in
  \cite{Note3}.~Except for $J_t=\int 1\circ\rmd x_\tau=x_t-x_0$ (blue
  line) we always choose the current defined with $U(\tau)=\nu(\tau)$ and density defined with $V(x,\tau)=-x\nu(\tau)$.}}
\label{Fg1}
\end{center}
\end{figure*}

To generalize to transients we use Eq.~\eqref{FPE} $\nabla_\x\cdot\f j(\x,\tau)=-\partial_\tau P(\x,\tau)$\blue{, integrate by parts twice (see \cite{Note3} for details), and define a second operationally accessible current
\begin{align}
\widetilde{J}_t\equiv\int_{\tau=0}^{\tau=t}\tau\partial_\tau\f U(\x_\tau,\tau)\cdot\circ\rmd\x_\tau\,,
\end{align}
to obtain
}
\begin{align}
\E{A_tJ_t^{\rm II}}=(t\partial_t-1)\E{J_t}-\E{\widetilde{J}_t}\,.
\label{At JIIt} \end{align}
Thus, we have expressed the correlation $\E{A_tJ_t^{\rm II}}$ in terms
of operationally accessible quantities. From this and Eq.~\eqref{TUR from ansatz},
the TUR for general initial conditions and general time-homogeneous Langevin dynamics Eq.~\eqref{SDE} reads
\begin{align}
  \Sigma_t \, \var(J_t)\ge 2   \left[t\partial_t\E{J_t}-\E{\widetilde{J}_t}\right]^2.
  \label{transient TUR}
\end{align}
The fact that the TUR for transient dynamics \eqref{transient TUR} follows from the
original TUR \eqref{TUR NESS} upon replacing $\E{J_t}\to t\partial_t
\E{J_t}$ is well known
\cite{Liu2020PRL,Pietzonka2017PRE} and was first derived in
continuous space in Ref.~\cite{Dechant2018JSMTE}.~However, the \blue{novel} correction
term $\E{\widetilde{J}_t}$ extends the validity of the TUR to currents
with an explicit time-dependence $\f U(\x,\tau)$.~We show below \blue{and in Fig.~\ref{Fg1}} that
this additional freedom in choosing $\f U$ is crucial for  saturating
the transient TUR under general conditions.~To highlight that
end-point derivative $t\partial_t$ 
and the correction term $\E{\widetilde{J}_t}$
are
strictly necessary we provide explicit counterexamples (see
\footnote{See Supplemental Material at [...] including Refs. \cite{Dieball2022NJP,Shiraishi2021JSP}.}).

We note that Eq.~\eqref{transient TUR} in one-dimensional space and
for additive noise can be deduced from restricting the result in
\cite{Koyuk2020PRL}, where an explicit time-dependence was introduced
via a speed parameter $v$, to a time-homogeneous drift, translated to
time-integrated 
currents, and noting that $v\partial_v U(x,v\tau)=\tau\partial_\tau
U(x,v\tau)$. The form without the speed parameter has the advantage
that the correction term $\E{\widetilde{J}_t}$ is accessible from a
single experiment while the $\partial_v$-correction requires
perturbing the speed of the experiment. However, the result in
\cite{Koyuk2020PRL} even holds for an explicitly time-dependent drift.

Notably, generalizing this proof to explicitly
time-dependent drift or diffusion, although probably possible, is
\emph{not} straightforward because it requires perturbing the
dynamics (see \cite{Koyuk2020PRL}), and therefore  all relevant
information is no longer contained in a single equation of motion.

\emph{TUR for densities.---}We define general, operationally
accessible densities (the term ``density'' is motivated by the analogy to ``current'' as e.g.\ in \cite{Dieball2022PRL,Dieball2022PRR,Touchette2018PA,Dieball_ARXIV_JPhysA_1})
\begin{align}
\rho_t&=\int_0^t V(\x_\tau,\tau)\rmd\tau\,,\nonumber\\
\widetilde{\rho}_t&\equiv\int_{\tau=0}^{\tau=t}\tau\partial_\tau V(\x_\tau,\tau)\rmd\tau\,.\label{definition density} 
\end{align}
Since in the proof above we did not use the explicit
form of $\mathcal U$, the density can be treated analogously to $J_t$
in Eq.~\eqref{def current Ito} by replacing $\mathcal U \to V$  and
omitting the $J_t^{\rm I}$-term. Analogously to Eqs.~\eqref{proof
  ansatz} and \eqref{At JIIt} we thus obtain
\begin{equation}
\E{A_t(\rho_t-\E{\rho_t})}=\E{A_t\rho_t}=(t\partial_t-1)\E{\rho_t}-\E{\widetilde{\rho}_t}\,,
\end{equation}
and analogously to Eq.~\eqref{TUR from ansatz} the
transient density-TUR
\begin{align}
\Sigma_t\var(\rho_t)\ge 2\left[(t\partial_t-1)\E{\rho_t}-\E{\widetilde{\rho}_t}\right]^2.
  \label{transient density TUR}  
\end{align}
Note that due to the absence of the $J_t^{\rm I}$-term, the right-hand side vanishes in steady-state systems.
As in the discussion of Eq.~\eqref{transient TUR} above,
Eq.~\eqref{transient density TUR} is in some sense contained in the
results of \cite{Koyuk2020PRL}. However, Eq.~\eqref{transient density
  TUR} allows for multidimensional space and multiplicative noise,
and does not require a variation in protocol speed. 

\emph{Improving TURs using correlations.---}It has been recently found
\cite{Dechant2021PRX} that the steady-state TUR can be eminently
improved, and even saturated arbitrarily far from equilibrium, by
considering correlations between currents and densities as defined in
Eq.~\eqref{definition density}. To re-derive this sharper version we rewrite 
 Eq.~\eqref{TUR from ansatz} for the observable $J_t-c\rho_t$ (the
 constant $c$ is in fact technically redundant since it can be absorbed in the definition of $\rho_t$)
\begin{align}
\frac{\Sigma_t}{2}\var(J_t-c\rho_t)&\ge\left[\E{J_t}+\E{A_t(J_t^{\rm II}-c\rho_t)}\right]^2\,.\label{CTUR from ansatz}
\end{align}
Note that
$\var(J_t-c\rho_t)=\var(J_t)+c^2\var(\rho_t)-2c\,\cov(J_t,\rho_t)$, where
cov denotes the covariance.
Using the optimal choice $c=\cov(J_t,\rho_t)/\var(\rho_t)$ and recalling that
for steady-state systems $\E{A_t(J_t^{\rm II}-c\rho_t)}=0$, 
Eq.~\eqref{CTUR from ansatz} becomes the NESS correlation-TUR in \cite{Dechant2021PRX}
\begin{align}
&\Sigma_t\,\var(J_t)\left[1-\chi^2_{J\rho}\right]\ge2 \E{J_t}^2,\nonumber\\
&\chi^2_{J\rho}\equiv\frac{\cov^2(J_t,\rho_t)}{\var(J_t)\var(\rho_t)}\,.
\label{CTUR NESS derived} 
\end{align}
Since $\chi^2_{J\rho}\in[0,1]$, Eq.~\eqref{CTUR NESS derived} is
sharper than Eq.~\eqref{TUR NESS} and, as proven in
\cite{Dechant2021PRX} and discussed below, for any steady-state system
there exist $J_t,\rho_t$ that saturate this inequality. 

Our approach allows to generalize this result to transient dynamics  by computing $\E{A_t(J_t^{\rm II}-c\rho_t)}$ as in Eq.~\eqref{At JIIt} to obtain from Eq.~\eqref{CTUR from ansatz} the \emph{generalized correlation-TUR}
\begin{align}
&\Sigma_t\,\var(J_t-c\rho_t)\ge \nonumber\\ & 2\left(t\partial_t\E{J_t}-\E{\widetilde{J}_t}-c\left[(t\partial_t-1)\E{\rho_t}-\E{\widetilde{\rho_t}}\right]\right)^2.
\label{transient CTUR}
\end{align}
One could again optimize the left-hand side over $c$ to obtain
$\var(J_t-c\rho_t)=\var(J_t)\left[1-\chi^2_{J\rho}\right]$. However,
since here the right-hand side also involves $c$ this may not be the
optimal choice. Thus, it is instead practical to keep $c$ general (or
absorb it into $\rho_t$). The generalized correlation-TUR
\eqref{transient CTUR} represents a \emph{novel result} that sharpens
the transient TUR in Eq.~\eqref{transient TUR}, and, as we show below \blue{and illustrate in Fig.~\ref{Fg1}},
even allows to generally saturate the TUR arbitrarily far from equilibrium. 

\emph{Saturation of TURs.---}For any choice $\f U$ in the definition
of $J_t$ in Eq.~\eqref{def current Strato}, the TUR allows to infer a lower bound on the time-accumulated dissipation $\Sigma_t$ from 
$\E{J_t}$ and $\var(J_t)$ 
\cite{Koyuk2021JPAMT,Horowitz2019NP,Gingrich2017JPAMT,Vu2020PRE,Manikandan2020PRL,Otsubo2020PRE,Li2019NC,Dieball2022PRL,Dieball2022PRR}.
The tighter the inequality, the more precise is the lower bound on
$\Sigma_t$. It is therefore important to understand when the
inequality becomes tight or even saturates, i.e.\ gives equality.

Due to the simplicity and directness of our proof, we can very well
discuss the tightness of the bound based on the \blue{single application of the Cauchy-Schwarz inequality. As elaborated in the Appendix, this approach reproduces, and extends beyond, numerous existing results on asymptotic and exact saturation of TURs.
Most importantly, choosing $\f U(\x,\tau)=c'[{\f
    j(\x_\tau,\tau)}/{P(\x_\tau,\tau)}]\cdot[2\f D(\x_\tau)]^{-1}$
with arbitrary $c'$ and $c\rho_t=J_t^{\rm II}$ (see Eq.~\eqref{def
  current Ito}) gives $J_t-c\rho_t=J_t^{\rm I}=c'A_t$ which in turn implies
equality in the Cauchy-Schwarz argument leading to the 
correlation-TURs Eqs.~\eqref{CTUR NESS derived} and \eqref{transient
  CTUR}. This directly implies exact saturation of the
correlation-TURs which was so far achieved only in the steady-state case \cite{Dechant2021PRX}. Our generalization of the correlation-TUR in Eq.~\eqref{transient CTUR} for transient systems 
therefore allows to saturate a TUR arbitrarily far from equilibrium for any $t$ and for general initial conditions and general time-homogeneous dynamics in Eq.~\eqref{SDE}.\\}
\indent \blue{\emph{Example.---}To illustrate the novel results in Eqs.~\eqref{transient TUR},
  \eqref{transient density TUR} and \eqref{transient CTUR} and the new
  insight into the saturation,  we provide an explicit example of
  transient dynamics in Fig.~\ref{Fg1}, that of a Brownian particle in
  a one-dimensional harmonic potential $\varphi(x,t)=a(x-x^0_t)^2/2D$
  displaced from $x^0_{\tau<0}=z$ to $x^0_{\tau\ge 0}=0$, see
  Fig.~\ref{Fg1}(a).~This setting, illustrated by the color gradient
  in Fig.~\ref{Fg1}a, can easily be realized experimentally using
  optical tweezers \cite{Crocker1994PRL,Curtis2002OC,Dasgupta2012OL}.   
The process features a
  Gaussian probability density $P(x,\tau)$ with constant variance $D/a$
  that moves with a space-independent velocity
  $\nu(\tau)=j(x,\tau)/P(x,\tau)=-az\exp(-a\tau)$ towards the
  equilibrium $\propto\exp(-ax^2/2D)$, see
  Fig.~\ref{Fg1}(b).\\
 \indent  To quantify the tightness of the respective TURs we inspect
  quality factors -- the ratio of the right- and left-hand side of the
  TUR -- shown in Fig.~\ref{Fg1}(c) as a function of the dimensionless
  quantity $at$.~ The blue line represents the transient TUR
  \eqref{transient TUR} for the current $J_t=x_t-x_0$ where
  $U(x,\tau)=1$. Since this $U$ does not feature explicit
  time-dependence the correction term $\widetilde{J}_t$ does not
  contribute and the transient TUR from the existing literature
  \cite{Dechant2018JSMTE} applies. The existing (as well as our)
  results allow varying the spatial dependence of $U$ but we refrain
  from considering this for simplicity and since it is not
  necessary for saturation (i.e.\ $\nu,D$ have no spatial dependence
  in our example). 
  Due to the novel correction term in Eq.~\eqref{transient TUR} we may choose a
  time-dependent $U$, and following our discussion of the saturation
  we choose for all following examples $J_t$ with
  $U(\tau)=c'\nu(\tau)/2D=\nu(\tau)$ (the prefactor $c'$
    is arbitrary as it cancels in quality factor) and the
  corresponding $\rho_t=J_t^{II}$, i.e.\ with $V(x,\tau)=\mathcal
  U(x,\tau)=-axU(\tau)$, see Eq.~\eqref{def current Ito}. For this
  choice we evaluate the transient current [Eq.~\eqref{transient TUR}] and
  density-TUR [Eq.~\eqref{transient density TUR}], see
  light gray and orange line in Fig.~\ref{Fg1}(c).~Moreover,
  we evaluate the novel generalized
  correlation-TUR~\eqref{transient CTUR} for $c=0.2$ (dark gray line),
  where we find that the current TUR is improved by considering
  correlations with the a density, and for $c=1$ (black line), where
  we find the expected saturation. This saturation means that the lower
  bound obtained for $\Sigma_t$ from this TUR is exactly
  $\Sigma_t$. Note that this exact saturation requires the knowledge of
  the details of the dynamics for the choice of $U,V$. However, even
  with very limited knowledge one can simply consider different guesses or
  approximations of the optimal $U,V$ and each guess will give a valid
  lower bound (given sufficient statistics).}\\
\indent \blue{\emph{Direct route for Markov jump processes.---}Beyond
  overdamped dynamics, one may employ the above direct approach for deriving
  TURs to Markov jump dynamics on a discrete state-space
$\mathcal N$ with jump-rates $(\rxy)_{x,y\in\mathcal N}$ and
  steady-state 
  distribution $(p_x)_{x\in\mathcal N}$. To illustrate this
  generalization, 
  we here provide the proof of the steady-state TUR~\eqref{TUR
    NESS}. Let $\tx$ denote the (random) time spent in state $x$
  and $\nxy$ the (random) number of jumps from $x$ to $y$ in the time
  interval $[0,t]$.  A general time-accumulated current in a jump
  process is defined with anti-symmetric prefactors $d_{xy}=-d_{yx}$
  as the double sum $J\equiv\sumxy d_{xy}\nxy$. The steady-state 
  dissipation in turn reads $\Sigma\equiv t\sumxy p_x\rxy\ln[{p_x\rxy}/{p_y\ryx}]$. Analogously to $A_t$ in Eq.~\eqref{def At} define
\begin{align}
A&\equiv\sumxy\frac{p_x\rxy-p_y\ryx}{p_x\rxy+p_y\ryx}(\nxy-\tx\rxy)\,.
\label{def A MJP} 
\end{align}
For this choice of $A$ one can check that $\E{A}=0,\ \E{A^2}\le\Sigma/2$, and $\E{AJ}=\E{J}$ (a ``direct'' proof as above follows by analogy of covariance properties of $\partial_t(\nxy-\tx\rxy)$ and $\bsig(\x_t)\rmd\f W_t$, see \cite{Note3} for details) which imply, via the Cauchy-Schwarz inequality, equivalently to Eqs.~\eqref{proof ansatz} and \eqref{TUR from ansatz} the steady-state TUR for Markov jump processes
\begin{align}
\E{A(J-\E{J})}=\E{J}\quad\Rightarrow\quad\frac{\Sigma}{2}\var(J)\ge\E{J}^2\,.\label{TUR MJP} 
\end{align}
A discussion of possible generalizations of this proof beyond
steady-state dynamics is given in \cite{Note3}.}\\
 \indent \emph{Conclusion.---}Using only stochastic calculus
and the well known Cauchy-Schwarz inequality we
proved various existing TURs directly from the Langevin equation.
This underscores the TUR as an inherent property of overdamped stochastic
equations of motion, analogous to quantum-mechanical uncertainty relations. 
Moreover, by including current-density correlations we derived a new
sharpened TUR for transient dynamics. Based on our simple and more direct
proof we were able to systematically explore conditions under which
TURs saturate. The new equality~\eqref{proof ansatz} is mathematically even stronger than
TUR~\eqref{TUR from ansatz}. Therefore it allows to derive further
bounds, e.g.\ by applying H\"older's instead of the Cauchy-Schwarz
inequality which, however, may not yield operationally
accessible quantities. Our approach may allow for generalizations to
systems with time-dependent driving (see e.g.\ \cite{Koyuk2020PRL}) which, however, are not expected
to follow anymore directly from a single equation of motion. \blue{The
  novel correction term for currents with explicit time dependence as
  well as
  the new transient correlation-TUR and its saturation are expected to
  equally apply to Markov jump processes by generalizing the approach
  illustrated in Eqs.~\eqref{def A MJP} and \eqref{TUR MJP}.\\}

\emph{Acknowledgments.---}We thank David Hartich for insightful suggestions and critical reading of the manuscript.\
Financial support from Studienstiftung des Deutschen Volkes (to
C.\ D.) and the German Research Foundation (DFG) through the Emmy
Noether Program GO 2762/1-2 (to A.\ G.) is gratefully acknowledged.\\
\\
\indent \blue{\emph{Appendix:~Saturation of TURs.---}}\blue{Thanks to the directness of our proof, we only need to discuss the tightness} based on the step from
Eq.~\eqref{proof ansatz} to Eq.~\eqref{TUR from ansatz} where we applied
the Cauchy-Schwarz inequality $\E{A_t(J_t-\E{J_t})}^2\le \E{A_t^2}\var(J_t)$
to the exact Eq.~\eqref{proof ansatz}.  Thus,
the closer $A_t$ and $J_t-\E{J_t}$ are to being linearly
dependent \blue{(recall that the Cauchy-Schwarz inequality
measures the angle $\varphi$ between two vectors $(\vec x\cdot\vec
y)^2 =\vec x^{\,2}\vec y^{\,2}\cos^2(\varphi)\le\vec x^{\,2}\vec
y^{\,2}$)},  the tighter the TUR, with saturation for
$J_t-\E{J_t}=c'A_t$ for some constant $c'$.  Therefore, the TUR is
expected to be tightest for the choice  $\f U(\x,\tau)=c'[{\f
    j(\x_\tau,\tau)}/{P(\x_\tau,\tau)}]\cdot[2\f D(\x_\tau)]^{-1}$ for
which $J_t^{\rm I}=c'A_t$ (see Eq.~\eqref{def current Ito}). Note that for
NESS this $\f U$ becomes time-independent
with $\f j_{\rm s}(\x)/P_{\rm s}(\x)$. This
choice is known to saturate the original TUR in Eq.~\eqref{TUR NESS}
in the near-equilibrium limit \cite{Pigolotti2017PRL}. 
However, since
the full $J_t=J_t^{\rm I}+J_t^{\rm II}$ current cannot be chosen to
exactly agree with $c'A_t$, equality is generally not reached. 

The original TUR~\eqref{TUR NESS} with this choice of $\f U(\x,\tau)$
was also found to saturate in the short-time limit $t\to 0$
\cite{Manikandan2020PRL,Otsubo2020PRE}. This result is in turn
reproduced with our approach by noting that $J_t^{\rm I}=c'A_t$ and
$\E{A_tJ_t^{\rm II}}=0$ give $\E{A_t(J_t-\E{J_t})}^2=\E{A_tJ_t^{\rm
    I}}^2=\E{A_t^2}\E{{J_t^{\rm I}}^2}$, and in the limit $t\to 0$ the
integrals in Eq.~\eqref{def current Ito} asymptotically scale like a
single time-step, such that $\E{{J_t^{\rm I}}^2}\sim(\f W_t-\f
W_0)^2\sim t$ dominates all $\sim t^{3/2},\,\sim t^2$ contributions in
$\var(J_t)$. In turn,
$\E{{J_t^{\rm I}}^2}\overset{t\to 0}{\to}\var(J_t)$
 which yields 
$\E{A_t(J_t-\E{J_t})}^2\overset{t\to 0}{\to}\E{A_t^2}\var(J_t)$.
   Thus, the Cauchy-Schwarz step
from the equality~\eqref{proof ansatz} to the inequality~\eqref{TUR from ansatz} saturates as $t\to 0$, in turn implying that
the TUR saturates. 

More recently it was also found that including correlations (see
Eq.~\eqref{CTUR NESS derived} and Ref.~\cite{Dechant2021PRX}) allows to
saturate a sharpened TUR for steady-state systems arbitrarily far from
equilibrium for any $t$, again for the same choice $\f U(\x,\tau)$ as
above.
Since our re-derivation of the NESS correlation-TUR in Eq.~\eqref{CTUR
  NESS derived} applied the Cauchy-Schwarz inequality to $A_t$ and
$J_t-c\rho_t$ we see that choosing $c\rho_t=J_t^{\rm II}$ yields
$J_t-c\rho_t=J_t^{\rm I}=c'A_t$, such that the application of the
Cauchy-Schwarz inequality becomes an equality. That is, the
correlation-TUR~\eqref{CTUR NESS derived} for this choice of $J_t$ and
$\rho_t$ is generally saturated. Notably, this powerful result follows
very naturally from the direct proof presented here. 

Our generalization of the correlation-TUR in Eq.~\eqref{transient
  CTUR} for transient systems even allows to saturate a TUR
(arbitrarily far from equilibrium for any $t$ and) for general initial
conditions and general time-homogeneous dynamics in 
Eq.~\eqref{SDE}.  This result is strong but obvious, since as for
the NESS correlation-TUR we can choose $J_t$ and $\rho_t$ such that
$J_t-c\rho_t=c'A_t$.  Note that it is here crucial that we allowed for
an explicit time-dependence in $\f U$ and $V$, i.e.\ that we found new
correction terms (terms with tilde in Eqs.~\eqref{transient
  TUR}, \eqref{transient density TUR} and \eqref{transient CTUR}).

\let\oldaddcontentsline\addcontentsline
\renewcommand{\addcontentsline}[3]{}
\bibliographystyle{apsrev4-2}
\bibliography{bib_TUR.bib}
\let\addcontentsline\oldaddcontentsline

\clearpage
\newpage
\onecolumngrid
\renewcommand{\thefigure}{S\arabic{figure}}
\renewcommand{\theequation}{S\arabic{equation}}
\setcounter{equation}{0}
\setcounter{figure}{0}
\setcounter{page}{1}
\setcounter{section}{0}

\begin{center}\textbf{Supplementary Material for:\\Direct Route to Thermodynamic Uncertainty Relations and Their Saturation}\\[0.2cm]
Cai Dieball and Alja\v{z} Godec\\
\emph{Mathematical bioPhysics Group, Max Planck Institute for Multidisciplinary Sciences, Am Fa\ss berg 11, 37077 G\"ottingen}\\[0.6cm]\end{center}
\begin{quotation}
\blue{In this Supplementary Material we provide a detailed analysis of the
quality (i.e.\ sharpness) of the distinct versions of thermodynamic uncertainty relation (TUR) applied to
the transient example shown in the Letter as well as counterexamples
underscoring the necessity of the novel versions of the TUR. Moreover,
we provide technical details on the direct proof of the TUR as well as
a perspective on the extension of the direct proof to Markov-jump
processes.}
\end{quotation}
\hypersetup{allcolors=black}\tableofcontents\hypersetup{allcolors=mylinkcolor}
\blue{\section{Quality of TURs for displaced harmonic trap}
In the following we derive the quality factors (i.e.\ the sharpness of the various TURs) shown in Fig.~1 in the Letter.}
Consider one-dimensional Brownian motion in a parabolic potential (i.e.\ a Langevin equation with linear force; known as the Ornstein-Uhlenbeck process) \cite{SM_Gardiner1985} with a Gaussian initial condition $x_0$ [we denote a a normal distribution by $\mathcal{N}({\rm mean,\,variance})$],
\begin{align}
\rmd x_\tau&=-ax_\tau\rmd \tau+\sqrt{2D}\rmd W_\tau\,,\qquad
x_0\sim\mathcal{N}(z,\sigma_0^2)\,.\label{SM OUP and IC}
\end{align}
Even though this process approaches an equilibrium steady state, for finite times it features transient dynamics if $x_0$ is not sampled from the steady-state distribution. For any Gaussian initial condition this process is Gaussian \cite{SM_Gardiner1985}. Therefore, the mean and the variance completely determine the distribution of $x_\tau$. \blue{The harmonic potential and the Gaussian initial condition can be realized experimentally by optical tweezers.}
The mean, variance and covariance are simply obtained as (see e.g.\ Appendix F in Ref.~\cite{SM_Dieball2022NJP})
\begin{align}
\E{x_\tau}&=z\rme^{-a\tau}\,,\nonumber\\
\var(x_\tau)&\equiv\E{x_\tau^2}-\E{x_\tau}^2=\frac{D}{a}\left(1-\rme^{-2a\tau}\right)+\sigma_0^2\rme^{-2a\tau}\,,
\nonumber\\
{\rm For\ }\tau\ge\tau'\colon\quad{\rm cov}(x_\tau,x_{\tau'})&\equiv\E{x_\tau x_{\tau'}}-\E{x_\tau}\E{x_{\tau'}}=\rme^{-a(\tau-\tau')}\var(x_{\tau'})\,.\label{SM covariance etc}
\end{align}
The Gaussian probability density $P(x,\tau)$ given the initial condition in Eq.~\eqref{SM OUP and IC} accordingly reads
\begin{align}
P(x,\tau)&=\sqrt{\frac{1}{2\pi\var(x_\tau)}}\exp\left[-\frac{(x-z\rme^{-a\tau})^2}{2\var(x_\tau)}\right]\,.\label{Gauss} 
\end{align}
The local mean velocity $\nu(x,\tau)\equiv j(x,\tau)/P(x,\tau)$ with current $j(x,\tau)\equiv(-ax-D\partial_x)P(x,\tau)$ reads
\begin{align}
\nu(x,\tau)=-ax+D(x-z\rme^{-a\tau})/\var(x_\tau)\,.\label{nu} 
\end{align}
For this example we consider the simple case $\sigma_0^2=D/a$,
i.e.\ we start in the steady-state variance (but as long as $z\ne 0$
not in the steady-state distribution\blue{; can be realized by
  equilibration with optical tweezers at $x=z$ at times $\tau<0$ and an
  equally stiff optical trap at position $x=0$ at times $\tau\ge0$}),
for which we obtain the simplified expressions
\begin{align}
\var(x_\tau)&=D/a\,,\nonumber\\
{\rm For\ }\tau\ge\tau'\colon\quad{\rm cov}(x_\tau,x_{\tau'})&=\rme^{-a(\tau-\tau')}D/a\,,\nonumber\\
\nu(x,\tau)&=-az\rme^{-a\tau}\,.\label{SM covariance etc simplified}
\end{align}
For this initial condition, $P(x,\tau)$ corresponds to a Gaussian distribution of constant variance with mean value $z\rme^{-a\tau}$ drifting from $z$ to $0$. Since only the mean changes (but the distribution around the mean remains invariant), the local mean velocity $\nu(x,\tau)$ is independent of $x$ [and in fact given by the velocity of the mean $\nu(x,\tau)=\partial_\tau\E{x_\tau}$]. This easily allows to compute the time-accumulated dissipation
\begin{align}
\Sigma_t=D^{-1}\int_0^t\rmd\tau\int\rmd x\E{\nu(x_\tau,\tau)}^2=\frac{a^2z^2}{D}\int_0^t\rmd\tau\rme^{-2a\tau}=\frac{az^2}{2D}(1-\rme^{-2at})\,.
\label{SM Sigma}
\end{align}
\blue{
Apart from $\Sigma_t$ the TURs contain first and second moments of (generalized) currents and densities that we derive for some examples below. Recall the definition of a generalized current (here in one-dimensional space)
\begin{align}
J_t&\equiv\int_{\tau=0}^{\tau=t}U(x_\tau,\tau)\circ\rmd x_\tau\,.
\end{align}
For simplicity (and since in our example the mean velocity $\nu(\tau)$ is space-independent) we consider only currents without explicit space dependence, i.e.\ only $U(x_\tau,\tau)=U(\tau)$. In this case there is no difference between the Stratonovich and It\^o interpretation of the integral, i.e.\ $J_t=\int_{\tau=0}^{\tau=t}U(\tau)\rmd x_\tau$.

\subsection{Current without explicit time-dependence}
For the simplest case of $U(x,\tau)=1$ we have the displacement current (denote this choice of current by $J_t^x$)
\begin{align}
J_t^x\equiv\int_{\tau=0}^{\tau=t}1\circ\rmd x_\tau=x_t-x_0\,.\label{SM Jx} 
\end{align}
From Eq.~\eqref{SM covariance etc simplified} we obtain
\begin{align}
\E{J_t}&=-z(1-\rme^{-at})\,,\qquad
t\partial_t\E{J_t}=-zat\rme^{-at}\,,\nonumber\\
\var(J_t)&=\var(x_t)+\var(x_0)-2{\rm cov}(x_t,x_0)=\frac{2D}{a}(1-\rme^{-at})\,.
\label{current 1st example}
\end{align}
Recalling the expression for the dissipation Eq.~\eqref{SM Sigma} the transient TUR $\Sigma_t\var(J_t)\ge2[t\partial_t\E{J_t}]^2$ in this example reads $\frac{az^2}{2D}(1-\rme^{-2at})\frac{2D}{a}(1-\rme^{-at})\ge 2(zat)^2\rme^{-2at}$ such that the quality factor $Q_x\in[0,1]$ (ratio of right-hand side and left-hand side; measures sharpness of the inequality) in this example becomes 
\begin{align}
Q_x\equiv\frac{2[t\partial_t\E{J_t}]^2}{\Sigma_t\var(J_t)}=\frac{2(at)^2\rme^{-2at}}{\left(1-\rme^{-2at}\right)\left(1-\rme^{-at}\right)}\,.\label{SM Qx}
\end{align}
Note that $Q_x$ is independent of $z$ and $D$ and only depends on the dimensionless quantity $at$. For large values $at\to\infty$ we have $Q_x\to 0$, see also Fig.~1 in the Letter.


\subsection{Currents with explicit time-dependence}
Saturating a TUR (i.e.\ obtaining the true dissipation as the lower bound inferred by the TUR) can 
be achieved for the 	transient correlation-TUR [Eq.~(22) in the Letter] by choosing $\f U(\x,\tau)=c'[{\f j(\x_\tau,\tau)}/{P(\x_\tau,\tau)}]\cdot[2\f D(\x_\tau)]^{-1}$,
and the corresponding density $\rho_t=\int_0^t\big[\f U(\x_\tau,\tau)\cdot\f
F(\x_\tau)+\nabla\cdot\left[\f D(\x_\tau)\f U(\x_\tau,\tau)\right] \big]\rmd\tau$. For this example, the respective current and density (denote this choice by superscript $\nu$) read (Stratonovich convention irrelevant since $\partial_x U(x,\tau)=0$ here)
\begin{align}
J_t^\nu&\equiv\frac{c'}{2D}\int_{\tau=0}^{\tau=t}\nu(\tau)\rmd x_\tau
\,,\qquad
\rho_t^\nu\equiv\frac{-ac'}{2D}\int_0^t x_\tau\nu(\tau)\rmd\tau
\,,
\end{align}
with $\nu(\tau)=-az\rme^{-a\tau}$, see Eq.~\eqref{SM covariance etc simplified}. The prefactor $c'$ will equally appear on both sides of the TURs and therefore not change the quality factors. One may set $c'=2D$ as done in the Letter, but here we keep it general. Plugging in $\rmd x_\tau$ from Eq.~\eqref{SM OUP and IC} we calculate
\begin{align}
\E{J_t^\nu}&=\frac{c'}{2D}\int_0^t\nu(\tau)\E{-ax_\tau}\rmd\tau
=\frac{-azc'}{2D}\int_0^t\nu(\tau)\rme^{-a\tau}\rmd\tau
=\frac{(az)^2c'}{2D}\int_0^t\rme^{-2a\tau}\rmd\tau=\frac{az^2c'}{4D}(1-\rme^{-2at})
\,,\nonumber\\
t\partial_t\E{J_t^\nu}&=t\frac{(az)^2c'}{2D}\rme^{-2at}\,,\qquad
\E{\rho_t^\nu}=\E{J_t^\nu}\,,\qquad
t\partial_t\E{\rho_t^\nu}=t\partial_t\E{J_t^\nu}\,.
\end{align}
The auxiliary current $\widetilde{J}_t\equiv\int_{\tau=0}^{\tau=t}\tau\partial_\tau U(x_\tau,\tau)\circ\rmd x_\tau$ and density $\widetilde{\rho}_t$ have due to $\tau\partial_\tau\nu(\tau)=-a\tau\nu(\tau)$ the mean
\begin{align}
\E{\widetilde{\rho}_t^\nu}=\E{\widetilde{J}_t^\nu}=\frac{-a(az)^2c'}{2D}\int_0^t\tau\rme^{-2a\tau}\rmd\tau
=\frac{-a(az)^2c'}{2D}\,\frac{1-\rme^{-2at}(1+2at)}{4a^2}
=\frac{-az^2c'}{8D}[1-(1+2at)\rme^{-2at}]\,.
\end{align}
For the variance split $J_t^\nu=J_t^{\rm I}+J_t^{\rm II}$ with
\begin{align}
J_t^{\rm I}&=\frac{c'}{2D}\int_{\tau=0}^{\tau=t}\nu(\tau)\sqrt{2D}\rmd W_\tau\,,
\qquad
J_t^{\rm II}=\rho_t^\nu
\,,
\end{align}
and compute
\begin{align}
\var(\rho_t^\nu)&=\var(J_t^{\rm II})=\frac{a^2c'^2}{4D^2}\Elr{\left(\int_0^t\nu(\tau)(\x_\tau-\E{\x_\tau})\rmd\tau\right)^2}\nonumber\\
&=\frac{a^2c'^2}{4D^2}2\int_0^t\rmd\tau\int_0^\tau\rmd\tau'\,\nu(\tau)\nu(\tau'){\rm cov}(x_\tau,x_{\tau'})\nonumber\\
&=\frac{a^3z^2c'^2}{2D}\int_0^t\rmd\tau\int_0^\tau\rmd\tau'\rme^{-a(\tau+\tau')}
\rme^{-a(\tau-\tau')}\nonumber\\
&=\frac{a^3z^2c'^2}{2D}\int_0^t\rmd\tau\tau\rme^{-2a\tau}
\nonumber\\
&=\frac{c'^2az^2}{8D}\left[1-(1+2at)\rme^{-2at}\right]\,.\label{SM var rho nu}
\end{align}
The cross terms are given by the non-trivial correlations of $\rmd W_\tau$ and $\rmd\tau'$ integrals exactly as Eq.~(13) in the Letter, in this example with $\mathcal U(x',\tau')=-ax'\nu(\tau')c'/(2D)$, such that
\begin{align}
\E{J_t^{\rm I}J_t^{\rm II}}
&=-\int_0^t\rmd\tau'\int\rmd x' \mathcal U( x',\tau')\int_0^t\rmd\tau\mathbbm1_{\tau<\tau'}\int\rmd x P( x',\tau'| x,\tau)\partial_x c'\nu(\tau)P( x,\tau)\nonumber\\
&=\frac{c'^2a}{2D}\int_0^t\rmd\tau'\int_0^{\tau'}\rmd\tau\nu(\tau)\nu(\tau')\int\rmd x' x'P( x',\tau'| x,\tau)\partial_x P( x,\tau)\,.
\end{align}
From Eq.~\eqref{Gauss} we get
\begin{align}
\partial_x P(x,\tau)=-(x-z\rme^{-a\tau})P(x,\tau)/\var(x_\tau)=-\frac{a}{D}(x-z\rme^{-a\tau})P(x,\tau)\,.
\end{align}
Hence using $z\rme^{-a\tau}=\E{x_\tau}$ and $\E{\E{x_{\tau'}}(x_\tau-\E{x_\tau})}=0$ we obtain
\begin{align}
\E{J_t^{\rm I}J_t^{\rm II}}
&=-\frac{c'^2a^2}{2D^2}\int_0^t\rmd\tau'\int_0^{\tau'}\rmd\tau\nu(\tau)\nu(\tau')\int\rmd x'(x-z\rme^{-a\tau})x'P( x',\tau';x,\tau)\nonumber\\
&=-\frac{c'^2a^2}{2D^2}\int_0^t\rmd\tau'\int_0^{\tau'}\rmd\tau\nu(\tau)\nu(\tau')\E{x_{\tau'}(x_\tau-\E{x_\tau})}\nonumber\\
&=-\frac{c'^2a^2}{2D^2}\int_0^t\rmd\tau'\int_0^{\tau'}\rmd\tau\nu(\tau)\nu(\tau')\cov(x_\tau,x_{\tau'})\,. 
\end{align}
Comparison with the third line in Eq.~\eqref{SM var rho nu} yields
\begin{align}
\E{J_t^{\rm I}J_t^{\rm II}}=-\var(\rho_t^\nu)\,.\label{SM E JI JII}
\end{align}
Therefore from $J_t^\nu=J_t^{\rm I}+J_t^{\rm II}$ with $\E{J_t^{\rm I}}=0$ we obtain
\begin{align}
\var(J_t^\nu)&=\E{{J_t^{\rm I}}^2}+\var(J_t^{\rm II})+2\E{J_t^{\rm I}J_t^{\rm II}}
=\frac{c'^2}{2}\Sigma_t-\var(\rho_t^\nu)
\overset{\rm Eqs.~\eqref{SM Sigma},\eqref{SM var rho nu}}{=}
\frac{c'^2az^2}{8D}\left[1+(2at-1)\rme^{-2at}\right]\,.\label{SM var time-dep} 
\end{align}
Since $\rho_t^\nu=J_t^{\rm II}$ and $\E{J_t^{\rm I}}=0$, Eq.~\eqref{SM E JI JII} implies
\begin{align}
\cov(J_t^\nu,\rho_t^\nu)=\cov(J_t^{\rm I}+J_t^{\rm II},J_t^{\rm II})=\E{J_t^{\rm I}J_t^{\rm II}}+\var(J_t^{\rm II})=0\,,\label{SM cov J rho}
\end{align}
which gives
\begin{align}
\var(J_t^\nu-c\rho_t^\nu)&=\var(J_t^\nu)+c^2\var(\rho_t^\nu)\nonumber\\
&=\frac{c'^2az^2}{8D}\Big(1+(2at-1)\rme^{-2at}+c^2\left[1-(1+2at)\rme^{-2at}\right]\Big)\,.\label{SM var J rho} 
\end{align}

We now have evaluated all expressions entering the various transient TURs for the current $J_t^\nu$ and density $\rho_t^\nu$. The quality factors (ratios of right- and left-hand side) $Q_J$ for the transient TUR $\Sigma_t\var(J_t)\ge 2[t\partial_t\E{J_t}-\E{\widetilde{J}_t}]^2$ [Eq.~(16) in the Letter], $Q_\rho$ for the transient density-TUR $\Sigma_t\var(\rho_t)\ge 2[(t\partial_t-1)\E{\rho_t}-\E{\widetilde{\rho}_t}]^2$ [Eq.~(19) in the Letter], and $Q_C(c)$ (function of $c$) for the transient correlation-TUR $\Sigma_t\var(J_t-c\rho_t)\ge 2(t\partial_t\E{J_t}-\E{\widetilde{J}_t}-c[(t\partial_t-1)\E{\rho_t}-\E{\widetilde{\rho_t}}])^2$ [Eq.~(22) in the Letter] after straightforward simplifications read
\begin{align}
Q_J&
=\frac{1-(1-2at)\rme^{-2at}}{2(1-\rme^{-2at})}\,,
\nonumber\\
Q_\rho&
=\frac{1-(1+2at)\rme^{-2at}}
{2(1-\rme^{-2at})}\,,
\nonumber\\
Q_C(c)
&=\frac{\left[(1+c)(1-\rme^{-2at})+(1-c)2at\rme^{-2at}\right]^2}
{2(1-\rme^{-2at})\left[(1+c^2)(1-\rme^{-2at})+(1-c^2)2at\rme^{-2at}\right]}\,,
\nonumber\\
Q_C(1)&
=1\,.\label{SM Q nu} 
\end{align}
These quality factors along with $Q_x$ in Eq.~\eqref{SM Qx} are depicted in Fig.~1 in the Letter. They only depend on the dimensionless quantity $at$ but not on other parameters of the process.

Opposed to $Q_x$ in Eq.~\eqref{SM Qx} these quality factors approach non-zero values as $at\to\infty$, namely $Q_J\to 1/2$, $Q_\rho\to 1/2$, $Q_C(c)\to (1+c)^2/2(1+c^2)$. Interestingly, for this special case the current and density quality factors add up to one, $Q_J+Q_\rho=1$. While $\lim_{at\to 0}Q_J=1$ [as expected in general for such choice of current (see ``saturation''-paragraph in the Letter)], we have $\lim_{at\to 0}Q_\rho=0$. This is because $at\to 0$ corresponds to the steady-state limit where the density-TUR becomes the trivial $\Sigma_t\var(\rho_t)\ge 0$.

Recall that $\cov(J_t^\nu,\rho_t^\nu)=0$ [see Eq.~\eqref{SM cov J rho}]. We now see that the choice $c=\cov(J_t,\rho_t)/\var(\rho_t)$, that is optimal in steady-state dynamics (see Letter), here gives $c=0$ although $c=1$ is optimal, see Eq.~\eqref{SM Q nu}. As mentioned in the Letter, this arises because $c=0$ only optimizes the left-hand side of the corelation-TUR [also seen from Eq.~\eqref{SM cov J rho}] but in the generalized (i.e.\ transient) correlation-TUR the right-hand side also depends on $c$.

\section{Counterexamples}
We showed that the TUR for transient dynamics [Eq.~(16) in the Letter] reads $\Sigma_t\var(J_t)\ge 2[t\partial_t\E{J_t}-\E{\widetilde{J}_t}]^2$ and thus contains the correction term $-\E{\widetilde{J}_t}$ which contributes if $\f U(\x,\tau)$ in the current $J_t\equiv\int_{\tau=0}^{\tau=t}U(\x_\tau,\tau)\circ\rmd x_\tau$ depends on time $\tau$. Based on the setting in the previous section (and Fig.~1 in the Letter) we here give an explicit counterexample for the TUR without the correction term, i.e.\ an example where $\Sigma_t\var(J_t)<2\left[t\partial_t\E{J_t}\right]^2$.} This shows that the correction term $-\E{\widetilde{J}_t}$ is indeed necessary, and that the result in Eq.~(16) in the Letter is valid for a broader class of systems than existing literature \cite{SM_Dechant2018JSMTE} by allowing explicit time-dependence in $\f U(\x_\tau,\tau)$. \blue{In addition we also provide a counterexample to show that the NESS TUR $\Sigma\var(J_t)\ge2\E{J_t}^2$ does not hold in this transient system.} 
\blue{Both counterexamples are shown in Fig.~\ref{FgSM} and the
  derivations of the respective terms are shown below.}

\begin{figure*}[ht!!]
\begin{center}
\includegraphics[width=.8\textwidth]{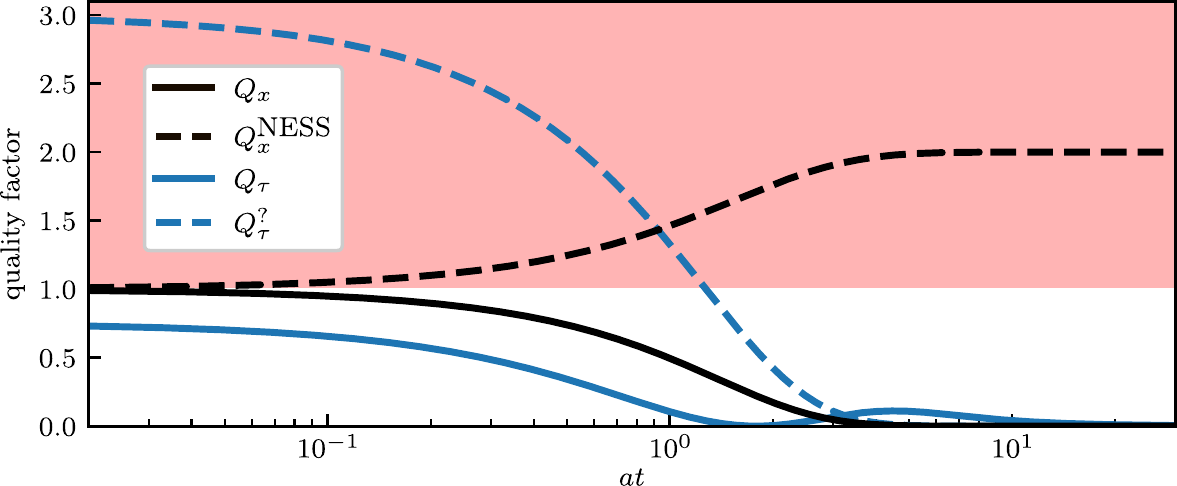}
\caption{\blue{Quality factors for the displaced harmonic trap for currents $J_t^x\equiv\int_{\tau=0}^{\tau=t}1\circ\rmd x_\tau=x_t-x_0$ (black) and $J_t^\tau\equiv\int_{\tau=0}^{\tau=t}\tau\circ\rmd x_\tau$ (blue). Quality factors $Q_{x,\tau}\equiv2[t\partial_t\E{J_t^{x,\tau}}-\E{\widetilde{J}_t^{x,\tau}}]^2/\Sigma_t\var(J_t^{x,\tau})\le1$, see Eqs.~\eqref{SM Qx} and \eqref{SM Qtau correct}, 
for the transient TUR 
[Eq.~(16) in the Letter] are shown by solid lines. The dashed black line $Q_x^{\rm NESS}\equiv2\E{J_t^x}^2/\Sigma_t\var(J_t^x)$, see Eq.~\eqref{SM Qx wrong}, 
 and dashed blue line $Q_\tau^?\equiv2[t\partial_t\E{J_t^\tau}]^2/\Sigma_t\var(J_t^\tau)$, see Eq.~\eqref{SM Qtau wrong}, intersect the forbidden region $Q>1$ (red background) which shows that the NESS TUR and the transient TUR without the correction $-\E{\widetilde{J}_t}$ term are not generally valid for transient systems.}
\label{FgSM}}
\end{center}
\end{figure*}

To find an example for which $2\left[t\partial_t\E{J_t}\right]^2>\Sigma_t\var(J_t)$, we note [recalling Eq.~(8) in the Letter, $\E{J_t}=\int_0^t\rmd\tau\int\rmd\x\f U(\x,\tau)\cdot\f j(\x,\tau)$] that the term $t\partial_t\E{J_t}=t\int\rmd\x\f U(\x,t)\cdot\f j(\x,t)$ only involves $\f U(\x,t)$ at the final time but not at any $\tau<t$. In contrast, $\Sigma_t$ is independent of the choice of $\f U$ and $\var(J_t)$ involves $\f U(\x,\tau)$ at all times. Therefore, examples for $2\left[t\partial_t\E{J_t}\right]^2>\Sigma_t\var(J_t)$ can be found by making $\f U(\x,t)$ large compared to $\f U(\x,\tau)$ at $\tau<t$. 

We now give an explicit example by choosing a current $J_t^\tau$ with linear time-dependence in $U(x,\tau)=\tau$ (here one-dimensional),
\begin{align}
J_t^\tau&\equiv\int_{\tau=0}^{\tau=t}\tau\circ\rmd x_\tau\,.
\end{align}
Note that due to $\partial_x U(x,\tau)=0$ there is no difference between Stratonovich and It\^o integration. We calculate
\begin{align}
\E{J_t^\tau}&=\int_0^t\tau\E{-ax_\tau}\rmd\tau=-az\int_0^t\tau\rme^{-a\tau}\rmd\tau=-\frac{z}{a}\left[1-\rme^{-at}(1+at)\right]\,,\nonumber\\
t\partial_t\E{J_t^\tau}&=\frac{zt}{a}\left[-a(1+at)+a\right]\rme^{-at}=-zat^2\rme^{-at}
\,.\label{SM Jtau mean}
\end{align}

\blue{For the variance compute analogously to the lines leading to Eq.~\eqref{SM var time-dep} [with cov from Eq.~\eqref{SM covariance etc simplified}]
\begin{align}
\var(J_t^\tau)&=2D\int_0^t\rmd\tau\,\tau^2-a^2\int_0^t\rmd\tau\int_0^t\rmd\tau'\,\tau\tau'{\rm cov}(x_\tau,x_{\tau'})\nonumber\\
&=2\frac{D}{a^3}\left(\frac{a^3t^3}{3}-\frac{1}{6}\left[a^2t^2(2at-3)-6\rme^{-at}(1+at)+6\right]\right)\nonumber\\
&=\frac{D}{a^3}\left[a^2t^2+2\rme^{-at}(1+at)-2\right]\,.
\end{align}
Since the dissipation $\Sigma_t$ does not depend on the choice of $J_t,U$, it is still given by Eq.~\eqref{SM Sigma}. The quality factor $Q_\tau^?$ for the TUR $\Sigma_t\var(J_t)\ge2\left[t\partial_t\E{J_t}\right]^2$ that would hold in the absence of explicit time-dependence in $U$ [Eq.~(16) in the Letter for $\E{\widetilde{J}_t}=0$; see also Ref.~\cite{SM_Dechant2018JSMTE}] reads
\begin{align}
Q_\tau^?
&=\frac{4a^4t^4\rme^{-2at}}{(1-\rme^{-2at})\left[a^2t^2+2\rme^{-at}(1+at)-2\right]}\,.\label{SM Qtau wrong}
\end{align}
Thus we see that as the other quality factors in Eqs.~\eqref{SM Qx} and \eqref{SM Q nu}, $Q_\tau^?$ only depends on the quantity $at$. In Fig.~\ref{FgSM} we see that $Q_\tau^?>1$ for small values of $at$ which breaks the TUR $\Sigma_t\var(J_t)\ge2\left[t\partial_t\E{J_t}\right]^2$, i.e.\ this example shows that the correction term $-\E{\widetilde{J}_t}$ in the TUR $\Sigma_t\var(J_t)\ge 2[t\partial_t\E{J_t}-\E{\widetilde{J}_t}]^2$ [Eq.~(16) in the Letter] is necessary for general validity for currents with explicitly time-dependent $\f U$.

For comparison we also give the correct quality factor $Q_\tau$ for the transient TUR [Eq.~(16) in the Letter] including the correction term $-\E{\widetilde{J}_t}$. Since $\tau\partial_\tau\tau=\tau$ we have $\widetilde{J}_t=J_t$ such that the correct quality factor using Eq.~\eqref{SM Jtau mean} reads
\begin{align}
Q_\tau
&=\frac{2\left(-zat^2\rme^{-at}+\frac{z}{a}\left[1-\rme^{-at}(1+at)\right]\right)^2}{\frac{az^2}{2D}(1-\rme^{-2at})\frac{D}{a^3}\left[a^2t^2+2\rme^{-at}(1+at)-2\right]}\nonumber\\
&=\frac{4\left[1-\rme^{-at}(1+at+a^2t^2)\right]^2}{(1-\rme^{-2at})\left[a^2t^2+2\rme^{-at}(1+at)-2\right]}\,.\label{SM Qtau correct}
\end{align}

To give a counterexample that shows that NESS TUR $\Sigma_t\var(J_t)\ge2\E{J_t}^2$ breaks down for this transient system, consider the current $J_t^x=x_t-x_0$ as in Eq.~\eqref{SM Jx}. The quality factor for this example follows from Eq.~\eqref{current 1st example},
\begin{align}
Q_x^{\rm NESS}
&=\frac{2z^2(1-\rme^{-at})^2}{\frac{az^2}{2D}(1-\rme^{-2at})\frac{2D}{a}(1-\rme^{-at})}
=\frac{2(1-\rme^{-at})}{1-\rme^{-2at}}\,.\label{SM Qx wrong}
\end{align}
Note that for any $y\in[0,1)$ the inequality $(1-y)^2>0$ implies $2(1-y)>1-y^2$. Thus we have for any value $at>0$ that $Q_x^{\rm NESS}>1$, see also Fig.~\ref{FgSM}. This provides (for any $at>0$) a counterexample against the NESS TUR, i.e.\ as expected the NESS TUR does not hold for transient dynamics.}

\textcolor{black}{
\section{Detailed derivation of $\E{A_t^2}=\Sigma_t/2$}
Recall the ``educated guess'' [Eq.~(9) in the Letter]
\begin{align}
A_t&\equiv\int_{\tau=0}^{\tau=t}\f a(\x_\tau,\tau)^T\rmd\f W_\tau\,,\nonumber\\
\f a(\x_\tau,\tau)^T&\equiv\frac{\f j(\x_\tau,\tau)^T}{P(\x_\tau,\tau)}[2\f D(\x_\tau)]^{-1}\bsig(\x_\tau)\,,\nonumber\\ 
\f a(\x_\tau,\tau)&=\bsig(\x_\tau)^T[2\f D(\x_\tau)]^{-1}\frac{\f j(\x_\tau,\tau)}{P(\x_\tau,\tau)}\,,\label{SM def At}
\end{align}
where $A_t$ and $P(\x_\tau,\tau)$ are scalars, $\f a(\x,\tau)$, $\f j(\x_\tau,\tau)$ and $\rmd\f W_\tau$ are vectors, and $2\f D(\x_\tau)=2\f D(\x_\tau)^T=\bsig(\x_\tau)\bsig(\x_\tau)^T$ are matrices (note that $([2\f D(\x_\tau)]^{-1})^T=[2\f D(\x_\tau)]^{-1}$).
Due to the ``delta-correlated'' noise property $\langle\rmd W_{\tau,i}\rmd
W_{\tau',j}\rangle=\delta(\tau-\tau')\delta_{ij}\rmd\tau\rmd\tau'$ the expectation of the square of the noise-integral $A_t$ is given by \cite{SM_Gardiner1985,SM_Pavliotis2014TiAM}
\begin{align}
\Elr{A_t^2}&=\int_0^t\Elr{\f a(\x_\tau,\tau)^T\f a(\x_\tau,\tau)}
\rmd\tau\nonumber\\
&=\int_0^t\Elr{\frac{\f j(\x_\tau,\tau)^T}{P(\x_\tau,\tau)}[2\f D(\x_\tau)]^{-1}\frac{\f j(\x_\tau,\tau)}{P(\x_\tau,\tau)}}
\rmd\tau.
\end{align}
Using that $\E{f(\x_\tau,\tau)}=\int\rmd\x P(\x,\tau)f(\x,\tau)$ and comparing to the definition of $\Sigma_t$ [Eq.~(5) in the Letter],
\begin{align}
\Sigma_t&=\int\rmd\x\int_{0}^{t}\frac{\f j(\x,\tau)^T\f D^{-1}(\x)\f j(\x,\tau)}{P(\x,\tau)}d\tau
\,,\label{SM_dissipation}
\end{align}
we immediately obtain $\E{A_t^2}=\Sigma_t/2$.
}

\blue{
\section{Detailed derivation of Eq.~(13)}
From the inequality $\Sigma_t\var(J_t)\ge2\left[\E{J_t}+\E{A_tJ_t^{\rm II}}\right]^2$ [Eq.~(12) in the Letter] we derive TURs by evaluating the expectation value (define notation $\f g(\x_\tau,\tau)\equiv\frac{\f j(\x_\tau,\tau)}{P(\x_\tau,\tau)}[2\f D(\x_\tau)]^{-1}$)
\begin{align}
\Elr{A_tJ_t^{\rm II}}&\equiv\Elr{\int_{\tau=0}^{\tau=t}\f g(\x_\tau,\tau)\cdot\bsig(\x_\tau)\rmd\f W_\tau\int_0^t \mathcal U(\x_{\tau'},\tau')\rmd\tau'}=\int_0^t\rmd\tau'\int_{\tau=0}^{\tau=t}\Elr{\f g(\x_\tau,\tau)\cdot\bsig(\x_\tau)\rmd\f W_\tau\mathcal U(\x_{\tau'},\tau')}\,.\label{SM E A JII} 
\end{align}
We use an approach from Refs.~\cite{SM_Dieball2022PRR,SM_Dieball2022JPA} to evaluate this expectation value. For convenience of the reader, we recall this approach here and apply it to this special case. 

First note that for times $\tau\ge\tau'$ this expectation value vanishes due
to the independence property of the Wiener process. However,
non-trivial contributions occur for $\tau<\tau'$ because the
probability density of $\x_{\tau'}$ depends on $\rmd\f W_\tau$. 
We want to express $\E{A_tJ_t^{\rm II}}$ in terms of integrals over the probability density $P(x,\tau)$ [that contains the information on the initial condition $P(x,0)$] and the conditional density $P(\x',\tau'|\x,\tau)$. This would be trivial in the absence of the noise increment $\bsig(\x_\tau)\rmd\f W_\tau$ by using $\E{\int_0^t V_1(\x_{\tau},\tau')\rmd\tau\int_\tau^t V_2(\x_{\tau'},\tau')\rmd\tau'}=\int\rmd\x\int\rmd\x'\int_0^t\rmd\tau\int_\tau^t\rmd\tau'V_1(\x,\tau)V_2(\x',\tau')P(\x',\tau'|\x,\tau)P(\x,\tau)$. The critical task is generalizing this to integration involving the noise increment.

For a given point $\x_\tau=\x$ we set $\beps\equiv\bsig(\x)\rmd\f W_\tau=\mathcal{O}(\sqrt{\rmd\tau})$ which has a Gaussian probability distribution $P(\beps)$ with zero mean and covariance matrix $2\f D (\x)\rmd\tau$. For a given $\beps$ the equation of motion in It\^o form implies a position increment $d\x_\tau(\x,\tau,\beps)=[\f F(\x)+\nabla\cdot \f D(\x)]\rmd\tau+\beps$. We now write the average in Eq.~\eqref{SM E A JII} as integrals over the probability density to be at points $\x,\x+d\x_\tau,\x'$ at times $\tau<\tau+d\tau<\tau'$, respectively, i.e.\ for $\tau<\tau'$
\begin{align}
\Elr{\f g(\x_\tau,\tau)\cdot\bsig(\x_\tau)\rmd\f W_\tau\mathcal U(\x_{\tau'},\tau')}=\!\int\!\rmd\x\!\int\!\rmd\x'\f g(\x,\tau)\cdot\beps\,\mathcal U(\x',\tau')P(\beps)P(\x',\tau'|\x+d\x_\tau(\x,\tau,\beps),\tau+d\tau)P(\x,\tau)\,.
\end{align}
Expanding in small $d\tau$ gives
\begin{align}
P(\x',\tau'|\x+d\x_\tau(\x,\tau,\beps),\tau+d\tau)=P(\x',\tau'|\x,\tau)+\rmd\x_\tau(\x,\tau,\beps)\cdot\nabla_\x P(\x',\tau'|\x,\tau)+\mathcal{O}(d\tau)\,.
\label{SM expand P}
\end{align}
By symmetry only the term of even power $\sim\beps^2$ in $\beps[\rmd\x_\tau(\x,\tau,\beps)\cdot\nabla_\x P(\x',\tau'|\x,\tau)]$ survives the integration over $P(\beps)$ and contributes according to the covariance matrix $2\f D(\x)\rmd\tau$. Therefore we arrive at (where $\mathbbm{1}_{\tau<\tau'}=1$ if $\tau<\tau'$ and $0$ otherwise)
\begin{align}
\E{\f g(\x_\tau,\tau)\cdot\bsig(\x_\tau)\rmd\f W_\tau \mathcal U(\x_{\tau'},\tau')}
=\mathbbm{1}_{\tau<\tau'}\rmd\tau\int\rmd\x\int\rmd\x' \mathcal U(\x',\tau')\f g(\x,\tau)2\f D(\x)P(\x,\tau)\cdot\nabla_\x P(\x',\tau'|\x,\tau)\,.
\end{align}
Now we perform an integration by parts in $\x$. The boundary terms
vanish in infinite space due to the vanishing of the probability
density at $\norm{\x}\to\infty$, in finite space with reflecting
(i.e.\ zero-flux) boundary conditions they vanish by the divergence theorem,
and in a finite system with periodic
boundary conditions the boundary terms cancel. Note that finite spatial domains
with reflecting boundary conditions are in essence already
contained in the infinite-space-case as the limit of a strongly confining potential. Using the symmetry $\f D^T(\x)=\f D(\x)$ we arrive at 
\begin{align}
\E{\f g(\x_\tau,\tau)\cdot\bsig(\x_\tau)\rmd\f W_\tau \mathcal U(\x_{\tau'},\tau')}
=-\mathbbm{1}_{\tau<\tau'}\rmd\tau\int\rmd\x\int\rmd\x' \mathcal U(\x',\tau')P(\x',\tau'|\x,\tau)\nabla_\x\cdot\left[2\f D(\x)\f g(\x,\tau)P(\x,\tau)\right]\,.
\end{align}
Plugging in the explicit form of $\f g(\x,\tau)$ we finally obtain
\begin{align}
\E{A_tJ_t^{\rm II}}&=-\int_0^t\rmd\tau'\int\rmd\x' \mathcal U(\x',\tau')\int_0^t\rmd\tau\mathbbm1_{\tau<\tau'}\int\rmd\x P(\x',\tau'|\x,\tau)\nabla_\x\cdot\f j(\x,\tau)\nonumber\\
&=-\int_0^t\rmd\tau'\int\rmd\x' \mathcal U(\x',\tau')\int_0^{\tau'}\rmd\tau\int\rmd\x P(\x',\tau'|\x,\tau)\nabla_\x\cdot\f j(\x,\tau)\,.
\label{SM lemma}
\end{align}

\section{Derivation of Eq.~(15)}
We here simplify Eq.~\eqref{SM lemma} for the case of transient
dynamics, i.e.\ where $\nabla\cdot\f j(\x,\tau)=-\partial_\tau
P(\x,\tau)$ does not vanish.} An integration by parts in $\tau$ with the boundary term ${-\int\rmd\x P(\x',\tau'|\x,0)P(\x,0)=-P(\x',\tau')}$ yields
\begin{align}
\E{A_tJ_t^{\rm II}}=\int_0^t\rmd\tau'\int\rmd\x'\mathcal U(\x',\tau')\left(-P(\x',\tau')-\int\rmd\x\int_0^t\rmd\tau P(\x,\tau)\partial_\tau\left[\mathbbm1_{\tau<\tau'}P(\x',\tau'|\x,\tau)\right]\right).
\end{align}
Note that the first term is $-\E{J_t^{\rm II}}$. Since we consider Markovian systems without explicit time-dependence of
$\f F$ and $\boldsymbol{\sigma}$, we have
$\partial_\tau P(\x',\tau'|\x,\tau)=\partial_\tau
P(\x',\tau'-\tau|\x)=-\partial_{\tau'}P(\x',\tau'-\tau|\x)=-\partial_{\tau'}
P(\x',\tau'|\x,\tau)$. Using moreover $\int\rmd\x P(\x',\tau'|\x,\tau)P(\x,\tau)=P(\x',\tau')$ and
$\int_0^t\rmd\tau\mathbbm1_{\tau<\tau'}=\tau'$ we obtain, upon
integrating 
by parts with the boundary term entering at $\tau'=t$, and recalling $\E{J_t^{\rm II}}=\E{J_t}$,
\begin{align}
\E{A_tJ_t^{\rm II}}&=-\E{J_t^{\rm II}}+\int\rmd\x'\int_0^t\rmd\tau'\mathcal U(\x',\tau'){\partial_{\tau'}}\left[\tau'P(\x',\tau')\right]\nonumber\\
&=(t\partial_t-1)\E{J_t}-\int\rmd\x'\int_0^t\rmd\tau'P(\x',\tau')\tau'\partial_{\tau'}\mathcal U(\x',\tau')\,.\label{SM intermediate result} 
\end{align}
In order to make Eq.~\eqref{SM intermediate result} operationally accessible we define a second current
\begin{align}
\widetilde{J}_t\equiv\int_{\tau=0}^{\tau=t}\tau\partial_\tau\f U(\x_\tau,\tau)\cdot\circ\rmd\x_\tau\,,
\end{align}
where $\E{\widetilde{J}_t}$ is analogously to Eqs.~(7) and (8) \blue{in the Letter} obtained via $\tau\partial_\tau\mathcal U$ such that \blue{from Eq.~\eqref{SM intermediate result} we obtain Eq.~(15) in the Letter, i.e.}
\begin{align}
\E{A_tJ_t^{\rm II}}=(t\partial_t-1)\E{J_t}-\E{\widetilde{J}_t}\,.
\end{align}

\blue{
\section{Extension to Markov jump processes}
We consider steady-state Markov jump dynamics. Recall from the Letter that currents are defined with anti-symmetric prefactors $d_{xy}=-d_{yx}$ as the double sum $J\equiv\sumxy d_{xy}\nxy$, and that the dissipation reads $\Sigma\equiv t\sumxy p_x\rxy\ln[{p_x\rxy}/{p_y\ryx}]$. Moreover, recall the definition [Eq.~(23) in the Letter]
\begin{align}
A&\equiv\sumxy Z_{xy}(\nxy-\tx\rxy)\nonumber\\
Z_{xy}&\equiv\frac{p_x\rxy-p_y\ryx}{p_x\rxy+p_y\ryx}
\,.
\label{SM def A Z} 
\end{align}
To complete the proof of the steady-state TUR as outlined in the Letter we need to show $\E{A}=0$, $\E{A^2}\le\Sigma/2$ and $\E{AJ}=\E{J}$. Proving these statements can be performed in complete analogy to the Cram{\'e}r-Rao proof of the steady-state TUR (the latter can e.g.\ be found in Ref.~\cite{SM_Shiraishi2021JSP}) by tilting the rates $\rxy(\theta)=\rxy\rme^{\theta Z_{xy}}$ and identifying $A=\partial_\theta\big|_{\theta=0}\ln\mathcal P_\theta$ where $\mathcal P_\theta$ is the tilted path measure. Note that $\E{\partial_\theta\ln\mathcal P_\theta}=\int\partial_\theta\mathcal P_\theta=\partial_\theta 1=0$ implies $\E{A}=0$. Using explicit properties of $Z_{xy}$ we have $\partial_\theta\E{J}_\theta=\E{J}_\theta$ \cite{SM_Shiraishi2021JSP} which implies 
$\E{J\partial_\theta \ln\mathcal P_\theta}_\theta=\partial_\theta\int J\mathcal P_\theta=\partial_\theta\E{J}_\theta=\E{J}_\theta$. Setting $\theta=0$ implies $\E{AJ}=\E{J}$. Evaluating the Fisher information $\mathcal I(\theta)$ at $\theta=0$ yields $\mathcal I(\theta=0)=\sumxy(p_x\rxy-p_y\ryx)^2/2(p_x\rxy+p_y\ryx)\le \Sigma/2$ \cite{SM_Shiraishi2021JSP} 
which due to $\E{A^2}=\E{(\partial_\theta\big|_{\theta=0}\ln\mathcal P_\theta)^2}=\mathcal I(\theta=0)$ concludes the proof.

Although we have completed the proof of the NESS TUR, this is by no means a ``direct'' proof, and it does not directly generalize in analogy to the generalizations performed in continuous space in the Letter. In order to give a genuinely ``direct'' proof in the sense that it completely follows from the equation of motion (i.e.\ from the properties of Markovian jumps and from the master equation), to generalize to arbitrary initial conditions, and to incorporate correlations of currents and densities, we consider a direct analogy to the continuous space approach presented in the Letter. We now give an outlook on this direct approach. Define $\hat c_{xy}(\tau)$ as the random variable representing a jump $x\to y$ at time $\tau$ such that $\nxy=\int_0^t\rmd\tau\hat c_{xy}(\tau)$ and $\hat{\mathbbm{1}}_x(\tau)$ as the random variable yielding $1$ if the state $x$ is occupied at $\tau$ (and $0$ otherwise) such that $\tx=\int_0^t\rmd\tau\hat{\mathbbm{1}}_x(\tau)$. Then define
\begin{align}
A\equiv\sumxy\int_0^t\rmd\tau\frac{p_x(\tau)\rxy-p_y(\tau)\ryx}{p_x(\tau)\rxy+p_y(\tau)\ryx}\left[\hat c_{xy}(\tau)-\rxy\hat{\mathbbm{1}}_x(\tau)\right]\,,\label{SM MJP direct A}
\end{align}
and split a current $J=\sumij d_{ij}\nij=\sumij d_{ij}\int_0^t\rmd\tau\hat c_{ij}(\tau)$ (where $d_{ij}=-d_{ji}$) into
\begin{align}
J&=J^{\rm I}+J^{\rm II}\,,
\nonumber\\
J^{\rm I}&=\sumij d_{ij}(\nij-\rij\ti)=\sumij d_{ij}\int_0^t\rmd\tau\left[\hat c_{ij}(\tau)-\rij\hat{\mathbbm{1}}_i(\tau)\right]\,,
\nonumber\\
J^{\rm II}&=\sumij d_{ij}\rij\ti=\sumij d_{ij}\rij\int_0^t\rmd\tau\hat{\mathbbm{1}}_i(\tau)\,.\label{SM MJP direct J}
\end{align}
Using this notation, one can directly show
that $\E{A}=0$, $\E{A^2}\le \Sigma/2$, $\E{AJ^{\rm I}}=\E{J}$ (analogous to proof in the Letter but instead of $\E{\sigma(x)\rmd W_\tau}=0$ and $\E{[\sigma(x)\rmd W_\tau]^2}=2D(x)\rmd \tau$ use that given $\hat{\mathbbm{1}}_x(\tau)=1$ the term $[\hat c_{xy}(\tau)-\rxy]\rmd\tau$ has zero mean and variance $\rxy\rmd\tau$). Thus, as in Eq.~(11) in the Letter we have
$\Sigma\var(J)\ge2\left[\E{J}+\E{AJ^{\rm II}}\right]^2$. Deriving the
different TURs then follows in analogy to the Letter by evaluating
$\E{AJ^{\rm II}}$ and accordingly introducing densities. All results,
including the discussion of the saturation, would then equally apply
to Markov jump processes, which will be addressed in the future.
}

\end{document}